\documentclass{sig-alternate-05-2015}

\usepackage{listings}
\usepackage{color}
\usepackage{balance}
\usepackage[T1]{fontenc}
\usepackage{microtype}
\usepackage{cite}
\usepackage{url}
\usepackage{tabularx}
\usepackage{booktabs}
\usepackage{multirow}
\usepackage{tikz}
\usepackage{amsmath}
\usepackage{ragged2e}
\usepackage{url} 
\usepackage{graphicx}
\usepackage{tikz}
\usepackage{tabularx}
\usepackage{booktabs}
\usepackage{hyperref}
\usepackage{caption}
\usepackage{array}
\usepackage{mdwmath}
\usepackage{mdwtab}
\usepackage{multirow}
\usepackage{longtable}
\usepackage{listings}
\usepackage{color}
\usepackage{balance}

\newcommand{\fakeparagraph}[1]{\smallskip\noindent\textbf{#1.}}

\newcommand*\circled[1]{\tikz[baseline=(char.base)]{
            \node[shape=circle,draw,inner sep=0.5pt, minimum size=0.5pt] (char) {#1};}}

\definecolor{name}{rgb}{0.5,0.5,0.5}
\definecolor{javared}{rgb}{0.6,0,0} % for strings
\definecolor{javagreen}{rgb}{0.25,0.5,0.35} % comments
\definecolor{javapurple}{rgb}{0.5,0,0.35} % keywords
\definecolor{javadocblue}{rgb}{0.25,0.35,0.75} % javadoc

\rmfamily
\DeclareSymbolFont{calletters}{OMS}{cmsy}{b}{n}
\DeclareSymbolFontAlphabet{\mathcal}{calletters}
\DeclareSymbolFont{rmletters}{OMS}{ptm}{m}{n}
\DeclareSymbolFontAlphabet{\mathrm}{rmletters}
\begin{document}

\setlength{\pdfpageheight}{\paperheight}
\setlength{\pdfpagewidth}{\paperwidth}
%\CopyrightYear{2016} 
%\setcopyright{acmcopyright}
%\conferenceinfo{SEsCPS'16,}{May 16 2016, Austin, TX, USA}
%\isbn{978-1-4503-4171-4/16/05}\acmPrice{\$15.00}
%\doi{http://dx.doi.org/10.1145/2897035.2897039}

\title{Evaluating a Development Framework for Engineering Internet of Things Applications}

\numberofauthors{3} %  in this sample file, there are a *total*
% of EIGHT authors. SIX appear on the 'first-page' (for formatting
% reasons) and the remaining two appear in the \additionalauthors section.
%
\author{
\alignauthor
Pankesh Patel\\
       \affaddr{ABB Corporate Research,}\\
			 \affaddr{Bangalore}\\
			 \affaddr{India}\\	
       \email{\normalsize{pankesh.patel@in.abb.com}}
% 2nd. author
\alignauthor
Tie Luo\\
       \affaddr{Institute for Infocomm Research, A*STAR}\\
			 \affaddr{Singapore}\\
       \email{\normalsize{luot@i2r.a-star.edu.sg}}
\and
\alignauthor Umesh Bellur\\
       \affaddr{Indian Institute of Technology Bombay }\\
			 \affaddr{India}\\		
       \email{\normalsize{umesh@cse.iitb.ac.in}}
}

\maketitle
\begin{abstract}
Application development in the Internet of Things (IoT) is challenging because it involves dealing with 
a wide range of related issues such as lack of separation of concerns in multiple layers and lack of high-level 
abstractions to address both the large scale and heterogeneity. Moreover, stakeholders involved in the application 
development have to address issues spanning to multiple life-cycles. Therefore, a critical 
challenge is to enable IoT application development with minimal effort from various stakeholders involved 
in the development process. 

Several approaches to tacking this challenge have been proposed in the fields of wireless sensor networks 
and ubiquitous and pervasive computing, regarded as precursors to the modern day of IoT. However, 
although existing approaches provide a wide range of features, stakeholders have specific 
application development requirements and choosing an appropriate approach requires 
thorough evaluations on different aspects.  To date, this aspect has been investigated to a limited
extend. In view of this, this paper provides an extensive set of evaluations based on our previous
work on IoT application development framework. Specifically, we evaluate our approach in terms of
(1) \emph{development effort}: the effort required to create a new application, (2) \emph{reusability}: 
the extend to which software artifacts can be reused during application development, 
 (3) \emph{expressiveness}: the characteristics of IoT applications that can be modeled 
using our approach, (4) \emph{memory metrics}: the amount of memory and storage a device needs to consume 
in order to run an application under our framework, and (5) \emph{comparison} of our 
approach with state of the art in IoT application development on various dimensions, which does 
not only provide a comprehensive view of state of the art, but also guides developers in
selecting an approach given application requirements in hand. We believe that the above 
different aspects provide the research community with insight into evaluating, selecting, and
developing useful IoT frameworks and applications.
\end{abstract}


\section{Introduction}
The Internet of Things~\cite[p.~6]{de2009internet} applications will involve interactions 
among extremely large numbers of heterogeneous devices, many of them directly interacting with 
their physical surroundings. Therefore, a critical challenge is to enable IoT application 
development with minimal effort from various stakeholders\footnote{Throughout this paper, we use the 
term \textbf{stakeholders} as used in software engineering to mean -- people, who are involved 
in the application development. Examples of stakeholders defined in \cite{softwareArchtaylor2010} 
are software designer, developer, domain expert, technologist, etc.} involved in the development process.
Similar challenges have already been addressed in the closely related fields of Wireless 
Sensor Networks (WSNs)~\cite[p.~11]{vasseur2010interconnecting} and ubiquitous and pervasive 
computing~\cite[p.~7]{vasseur2010interconnecting}, regarded as precursors to the modern day IoT. 
While the main challenge in the former is the \emph{large scale} -- hundreds to thousands of largely similar devices,  
the primary concern in the latter  has been the {\em heterogeneity} of devices and the major role that the  
user's own interaction with these devices plays in these systems (cf. the classic ``smart home'' scenario 
where a user controls lights and receives notifications from  his  refrigerator and toaster.).  
It is the goal of our work to enable the development of such applications. In the following, we discuss 
one of such applications. 

\subsection{Application example}\label{sec:appexample}
\begin{figure*}[htp]
\centering
\includegraphics[width=0.60\textwidth]{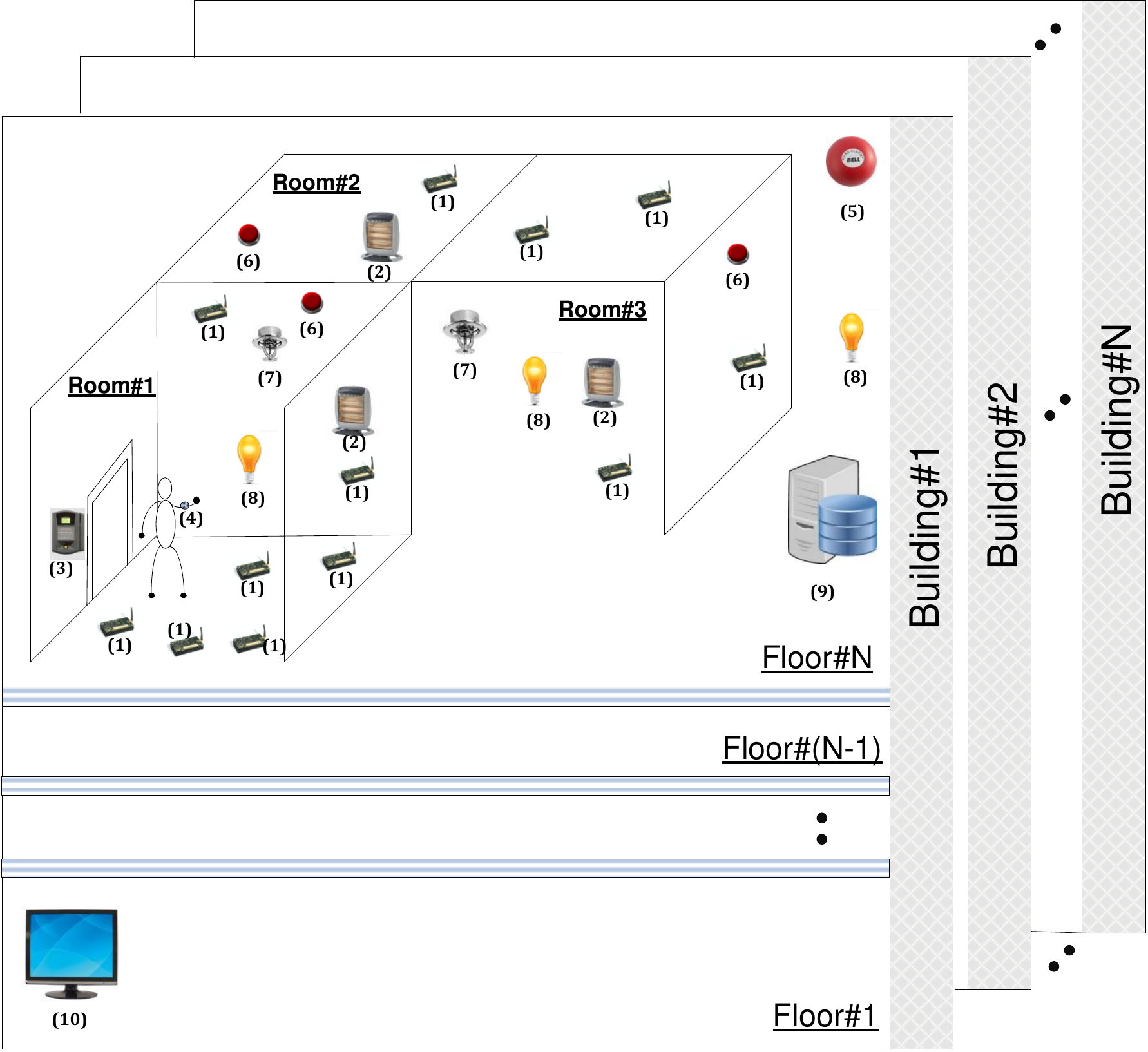}
\caption{A cluster of multi-floored buildings with deployed devices with (1)~temperature sensor, 
 (2)~heater, (3)~badge reader, (4)~badge, (5)~alarm, (6)~smoke detector, (7)~sprinkler, (8)~light, (9)~data storage, and 
 (10)~monitor.}
\label{fig:casestudybuilding}
\end{figure*}

To illustrate the characteristics of IoT applications, we consider the building automation 
domain~\cite[p.~361]{vasseur2010interconnecting}. This building system might 
consist of several buildings, with each building in turn consisting of one or more floors, 
each with several rooms that have a large number of heterogeneous devices equipped with sensors, actuators, 
storage, and user interfaces. Figure~\ref{fig:casestudybuilding} describes such a building automation 
system. Many applications can be developed using the in-built devices, one of 
which we discuss below.

\fakeparagraph{Personalized HVAC application}
This application aims to regulate temperature for workers'
productivity and personal comfort. To accommodate the workers' preference in the room, 
a database is used to keep the profile of each worker, including his preferred temperature level. 
A badge reader in the room detects the worker's entry event and queries the database 
for the worker's preference. Based on this, the thresholds used by the room's devices 
are updated. To reduce electricity waste when a person leaves the room, detected by 
badge disappeared event, heating is automatically set to the lowest level or 
according to the building's energy target. 

\subsection{IoT application development challenges}\label{sec:challenges}
This section reviews the application development challenges as gleaned from our analysis of 
applications such as the one discussed above. These challenges are as follows: 

\fakeparagraph{\emph{Heterogeneity}} 
IoT applications execute on a network consisting of heterogeneous 
devices  in terms of types~(e.g., sensing, actuating, storage, and user interface devices), interaction 
modes~(e.g. periodic~\cite{programmingWSNmottola2011}, publish/subscribe~\cite{eugster2003many}, 
request/response~\cite{berson1996client}, command~\cite{andrews1991paradigms}), 
different platforms~(e.g., Android mobile OS, Java SE on laptops), as well as different runtime system~(e.g., MQTT, DDS). 
The heterogeneity largely spreads into the application code and makes the portability of code to a different deployment difficult.  

\fakeparagraph{\emph{Large number of devices}} 
IoT applications may execute on distributed systems consisting of hundreds to 
thousands of devices, involving the coordination of their activities. Requiring the ability of 
reasoning at such levels of scale is impractical in general, as has been largely the view 
in the WSN community. 

\fakeparagraph{\emph{Different life cycle phases}} 
Like any other application development, the IoT application development is attributed 
to different-life cycle phases~\cite{Sommerville10}. At the \textbf{design phase}, the application logic 
has to be analyzed and separated into a set of distributed tasks for the underlying network consisting of a large 
number of heterogeneous entities. At \textbf{implementation phase}, the tasks have to be implemented 
for the specific platform to a device. At the \textbf{deployment phase}, the application logic has to be 
deployed onto a large number of devices.  Moreover, stakeholders have to keep 
in mind \textbf{evolution} issues both in the development (change in functionality of an application such as  the smart  
building application is  extended by including fire detection functionality) and deployment phase 
(e.g. adding/removing devices in deployment scenarios such as  more temperature sensors are added  
to sense accurate temperature values in the building). Manual effort in all above phases for 
a large number of heterogeneous devices is a time-consuming and error-prone process.

%As evident above, \emph{an important challenge that needs to be addressed in the IoT is 
%to enable the rapid development of IoT applications with minimal effort by the various 
%stakeholders involved in the process}.
\subsection{Contributions}

As evident above, an important challenge that needs to be addressed in the IoT is 
to enable the IoT application development with minimal effort by the various 
stakeholders. To address this challenge, many approaches have been proposed.
Although existing approaches provide a wide range of features, stakeholders have 
specific application development requirements and choosing an appropriate approach requires thorough 
evaluations on different aspects. To date, this aspect has been investigated to a limited extend for 
IoT applications, given heterogeneity at different life-cycle phases. Largely, the  metric 
used to asses the stakeholders' productivity in existing approaches is the number of lines of 
code~\cite[p.~45]{programmingWSNmottola2011}~\cite[p.~22]{programmingSurveysugihara2008} 
and it provides a little guidance to stakeholders to select an appropriate 
approach given application requirements in hand.  In view of these, this paper provides 
an extensive set of evaluations based on previous work~\cite{patelhalscube, patel-icse14, patel201562} 
on IoT application development framework, we call it as IoTSuite\footnote{An open source version, targeting on 
Android- and JavaSE -enabled devices and MQTT middleware, is available on: 
\url{https://github.com/pankeshlinux/IoTSuite/wiki}}. We believe that the following reported assessments not only be
helpful for our research, but might provide the research community with insights into evaluating, selecting, 
and developing useful IoT frameworks and applications.

\begin{itemize}
	
		\item \textbf{\emph{Development effort}}: In order to measure effort to develop an application 
		using our approach, we evaluate a percentage 	of a total number of lines 	of code generated 
		by our approach~(Section~\ref{sec:developmenteffort}). 	Moreover, 	we measure development effort 
		for an application that involves a large number of devices~(Section~\ref{sec:large}). 
		The results of both these experiments conclude that there is a drastic reduction 
		in development effort.

	\item \textbf{\emph{Reusability}}:  We evaluate reuse of specifications
	and implementations across applications using our approach. We consider
	different scenarios and demonstrate the development effort using our 
	approach to handle them. The results of these experiments conclude that 
	there is a drastic reduction in development effort for subsequent 
	application development~(Section~\ref{sec:reusability}).
	
	\item \textbf{\emph{Expressiveness}}: We evaluate the scope of our approach.  More specifically, we 
	answer the question: \emph{What are the characteristics of IoT applications that can be 
	modeled by our approach?}. We presents various \emph{characteristics} of IoT applications.
	Then, we map the representative IoT applications into identified characteristics using 
	our approach~(Section~\ref{sec:expressiveness}). Such an explicit evaluation may not only 
	be helpful as a framework for discussing coordinated research~(e.g., avoiding duplicate work) 
	in the IoT field, but may provide a basis for the development of software framework 
	to meet different IoT application requirements. 
	
	%In order to answer this question, we begin by presenting various 
	%\emph{characteristics} of IoT applications~(Section~\ref{abc}). Then, we map the representative IoT applications  
	%into identified characteristics using our approach~(Section~\ref{sec:expressiveness}).

	\item \textbf{\emph{Memory metrics}}: We measure the \emph{code size} and \emph{memory consumption} 
	of device-specific code that 	is deployed on devices. These two metrics are important 
	because they give approximate indication of the amount of memory a device need 
	to run an application~(Section~\ref{sec:codemetrics}).
	
	\item \textbf{\emph{Comparison with state of the art}}:  We begin by various \emph{dimensions} 
	that 	characterize IoT application development approaches. Then, we map existing approaches back to 
	the dimensions, which does not only provide a comprehensive view of state of the art, but also
	guides developers in selecting an approach given application requirements in 
	hand~(Section~\ref{sec:comparison}).

\end{itemize}

\fakeparagraph{Outline} The remainder of this paper is organized as follows: Section~\ref{sec:approach} summarizes our 
IoT application development process. This includes a brief on modeling languages and automation techniques. 
Section~\ref{sec:evaluation} evaluates the development framework in a quantitative manner. 
 Section~\ref{chapt:conclusionandfuturework} concludes this paper.

\section{IoT application development process}\label{sec:approach}

To provide the reader necessary background, this section summarizes our IoT application 
development process~(a complete tour is available in our previous publication~\cite{patel201562}), 
illustrated in Figure~\ref{fig:devcycle}. It separates IoT application development into different 
concerns and integrates a set of high-level languages to specify them.  
It is supported by compiler, mapper, and linker modules at various phases of IoT application 
development process to provide automation. Stakeholders carry out 
the following steps in order to develop an IoT application using our approach.

\subsection{Domain concern}
This concern is related to concepts that are specific to a domain (e.g., building automation, transport) 
of an IoT application. It consists of the following steps:

\fakeparagraph{\emph{Specifying domain vocabulary}} 
The domain expert specifies a domain vocabulary~(step~\circled{1} in Figure~\ref{fig:devcycle}) using
vocabulary language~(VL). The vocabulary specification 
includes concepts specific to a target application domain. For example, the building automation domain 
is reasoned in terms of rooms and floors, while the transport domain is expressed in terms 
of highway sectors. Furthermore, the vocabulary includes specification of resources, which are responsible for 
interacting with entities of interest~(EoI). This includes sensors~(sense EoI), actuators~(control EoI), 
and storage~(store information about EoI).  

%The development framework provides an editor support to write high-level specifications~(i.e, vocabulary, architecture, 
%and deployment) with the facilities of syntax coloring and syntax error reporting. We use 
%Xtext\footnote{\url{http://www.eclipse.org/Xtext/}} for a full fledged editor support. The Xtext 
%is a framework for a development of domain-specific languages and provides an editor support by writing Xtext grammar.

\fakeparagraph{\emph{Compiling vocabulary specification}}
Leveraging the vocabulary, the development framewotk generates~(step~\circled{2} in Figure~\ref{fig:devcycle}): 
(1) a vocabulary framework to aid the device developer, (2) a customized architecture grammar according to the vocabulary 
to aid the software designer, and (3) a customized deployment grammar according to the vocabulary 
to aid the network manager. The key advantage of this customization  is that 
domain-specific concepts defined in the vocabulary are made available to other stakeholders and can 
be reused across applications of the same application domain.

\subsection{Functional concern}
\begin{figure*}[!ht]
\centering
\includegraphics[width=0.7\linewidth]{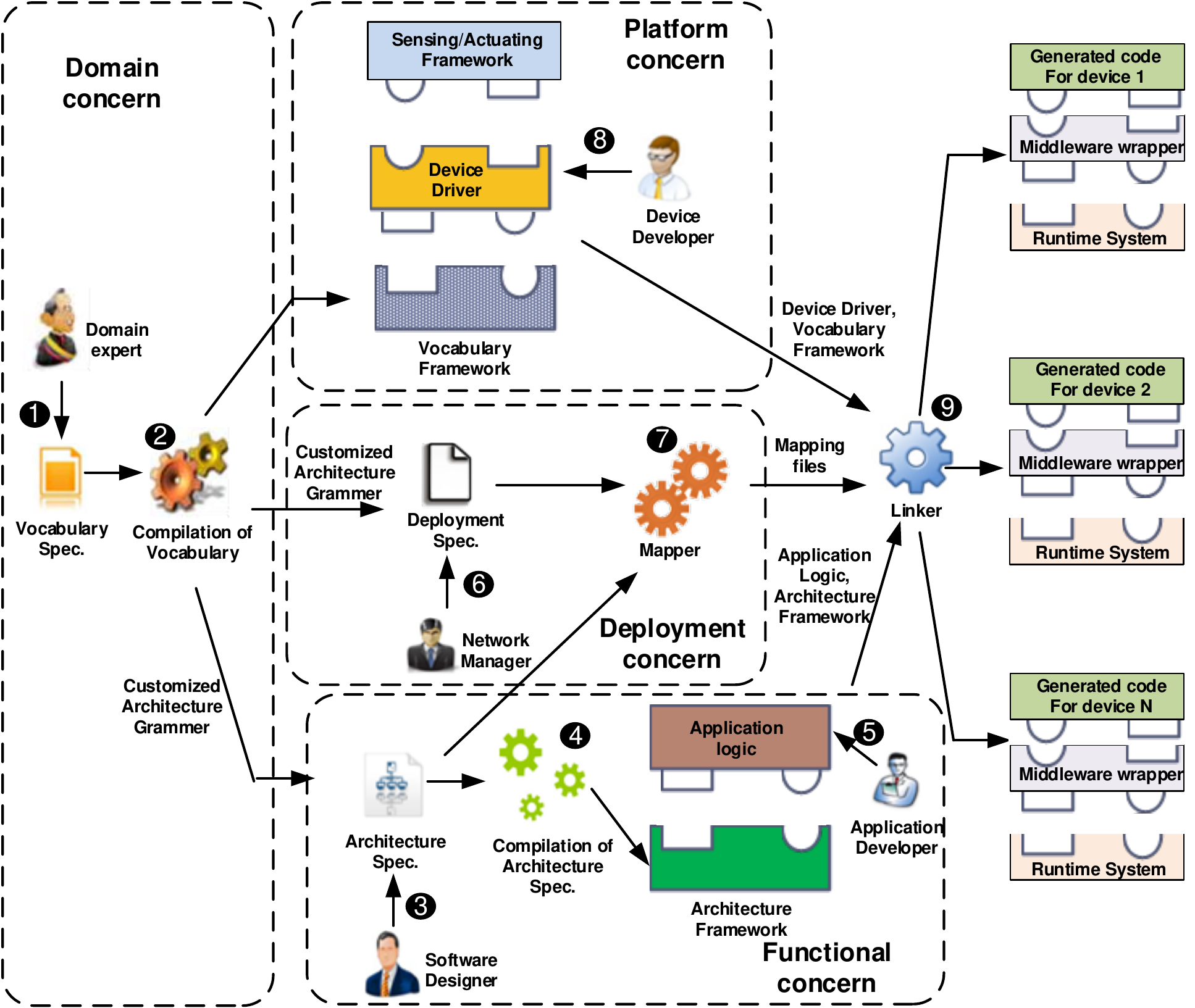}
\caption{IoT application development: overall process}
\label{fig:devcycle}
\end{figure*}

This concern is related to concepts that are specific to functionality of an IoT application. 
An example of a functionality is to open a window when an average temperature value of a
room is greater than $30^{\circ} C$. It consists of the following steps:

\fakeparagraph{\emph{Specifying application architecture}}
Using a customized architecture grammar, the software designer 
specifies  an application architecture~(step~\circled{3} in Figure~\ref{fig:devcycle}) using architecture language~(AL).
He specifies computational services and interactions with other
components. Computational services are fueled by sensors and storage (defined in the vocabulary). 
They process inputs data and take appropriate decisions by triggering actuators (defined in the
vocabulary specification).

\fakeparagraph{\emph{Compiling architecture specification}}  
The development framework leverages an architecture specification to support 
the application developer~(step~\circled{4} in Figure~\ref{fig:devcycle}). To describe the application logic of
each computational service, the application developer is provided an
architecture  framework, pre-configured according to the
architecture specification of an application, 
an approach similar to the one discussed in~\cite{towardscassou2011, generativecassou2009}. 

\fakeparagraph{\emph{Implementing application logic}} 
To describe the application logic of each computational service, the application 
developer leverages a generated architecture framework~(step \circled{5} in Figure~\ref{fig:devcycle}). It contains abstract 
classes\footnote{We assume that  the application developer uses an object-oriented 
language.}, corresponding to  each  computational service, 
that hide interaction details with other software components  and allow the application developer to 
focus only  on application logic. The application developer implements only the abstract 
methods of generated abstract classes.

\subsection{Deployment concern}
The concepts that fall into this concern describe information about device and its properties, 
placed in deployment scenario. It consists of the following steps:

\fakeparagraph{\emph{Specifying target deployment}} 
Using a customized deployment grammar, the network manager describes a deployment specification~(step~\circled{6} 
in Figure~\ref{fig:devcycle}) using deployment language~(DL). The deployment specification  includes the details of each 
device, including its regions (in terms of values of the regions defined in the vocabulary), resources hosted by devices (a subset of those defined 
in the vocabulary), and the type of the device. Ideally, the same IoT application could be deployed on different target deployments~(e.g., the same inventory tracking 
application can be deployed in different warehouses). This requirement is dictated by separating  a deployment specification  from other specifications.

\fakeparagraph{\emph{Mapping}}
The mapper produces a mapping from a set of computational services to a set of devices~(step~\circled{7} in Figure~\ref{fig:devcycle}). It takes as input a set of placement rules of computational services from an architecture 
specification and a set of devices defined in a deployment specification. The mapper decides where each computational service will be deployed. The current version of algorithm~\cite{patel-thesis14} selects devices randomly and allocates computational services to the selected devices. A mapping algorithm aware of heterogeneity, associated 
with devices of a target deployment, is a part of our future work.

\subsection{Platform concern}
This concern specifies the concepts that fall into this  are computer programs that act as a translator between a hardware device and an application. It consists of the following steps:

\fakeparagraph{\emph{Implementing device drivers}}
Leveraging the vocabulary, IoTSuite generate a vocabulary framework to aid the device developer~(step~\circled{8} in Figure~\ref{fig:devcycle}). The vocabulary framework 
contains a set of {\em interfaces} and {\em concrete classes} corresponding to resources defined in the vocabulary. The concrete classes contain concrete methods for interacting 
with other software components and platform-specific device drivers. We have integrated existing open-source sensing frameworks\footnote{\url{http://www.funf.org/}} for Android devices. 
So, the device developer has to only implement interfaces, connecting integrated sensing framework and generated vocabulary framework.

\subsection{Linking}
The linker combines and packs code generated by various stages into packages that can be 
deployed on devices~(step~\circled{9} in Figure~\ref{fig:devcycle}).  This  stage supports the application deployment 
phase by producing device-specific code to result in a distributed software system collaboratively 
hosted by individual devices, thus providing automation at  the deployment phase. 

The final output after linking is composed of three parts: (1) a \emph{runtime-system} runs on each individual device and provides a support for executing distributed tasks, (2) a \emph{device specific
code} generated by the linker module, and (3) a \emph{wrapper} separates generated code from the linker module and underlying runtime system by implementing interfaces. The main advantage of separating wrapper and runtime system is that
developers has to implement given interfaces, discussed in \cite{soukaras2015iotsuite}, in order to integrate a new runtime system. The current implementation of development framework implements the MQTT\footnote{\url{http://mqtt.org/}} 
and iBICOOP~\cite{bennaceur2009ibicoop}  runtime system, which enables interactions among Android devices and  JavaSE enabled devices.

\subsection{Evolution}
Evolution is an important aspect in IoT application development where sensors, actuators, and computational services are added, removed, or extended. 
To deal with these changes, we separate IoT application development into different concerns and allow an iterative development for these concerns. 
This iterative development requires only a change in evolved specification and reusing dependent specifications/implementation in compilation process, thus reducing effort 
to handle evolution, similar to the work in~\cite{towardscassou2011}. 

\section{Evaluation}\label{sec:evaluation}
The goal of this section is to describe how well the proposed approach addresses our aim.  Unfortunately, quality measures are not well-defined and they do not provide
a clear procedural method to evaluate development approaches in general. We established a set of measures and metrics that are vital for the productivity of stakeholders.
The set of measures is non-exhaustive. However, they reflect principal quantitative advantages that our approach provides to stakeholders involved in IoT application 
development. We evaluate our approach in terms of \emph{development effort}, \emph{reusability},\emph{expressiveness}, \emph{memory metrics}, and {comparison with state of the art}. 

\subsection{Development effort}\label{sec:developmenteffort}
In order to measure effort to develop an application using our approach, we evaluate a percentage of a total number of lines of code generated by our approach.
This section is organized as follows: Section~\ref{sec:appdevelopmenteffort} describes two applications to evaluate the development effort. Section~\ref{sec:resultdevelopmenteffort} shows results we obtained that 
indicates the effort required to create these two applications.

\subsubsection{Applications for evaluating development effort}\label{sec:appdevelopmenteffort} 
To evaluate development effort using our approach, we consider representative IoT applications.
Figure~\ref{fig:buildingautomationdomain} describes the building automation domain with various devices. Many applications can be developed using these devices.
We describes two applications: (1) {\emph{a personalized HVAC application}~(discussed in Section~\ref{sec:appexample}) 
and (2) \emph{a fire detection application}. It aims to detect fire by analyzing data from smoke and temperature sensors. Figure~\ref{fig:personalizedHVACandFiredetection} shows a layered architecture 
of both applications. In the fire detection application, a fire state is computed based on a current average temperature  value and smoke presence. Finally, the fire controller decides whether 
alarms should be activated or not. 

%We first give a general idea about the domain, followed by the details of the smart building application.

\begin{figure*}[!ht]
\centering
\includegraphics[width=0.7\linewidth]{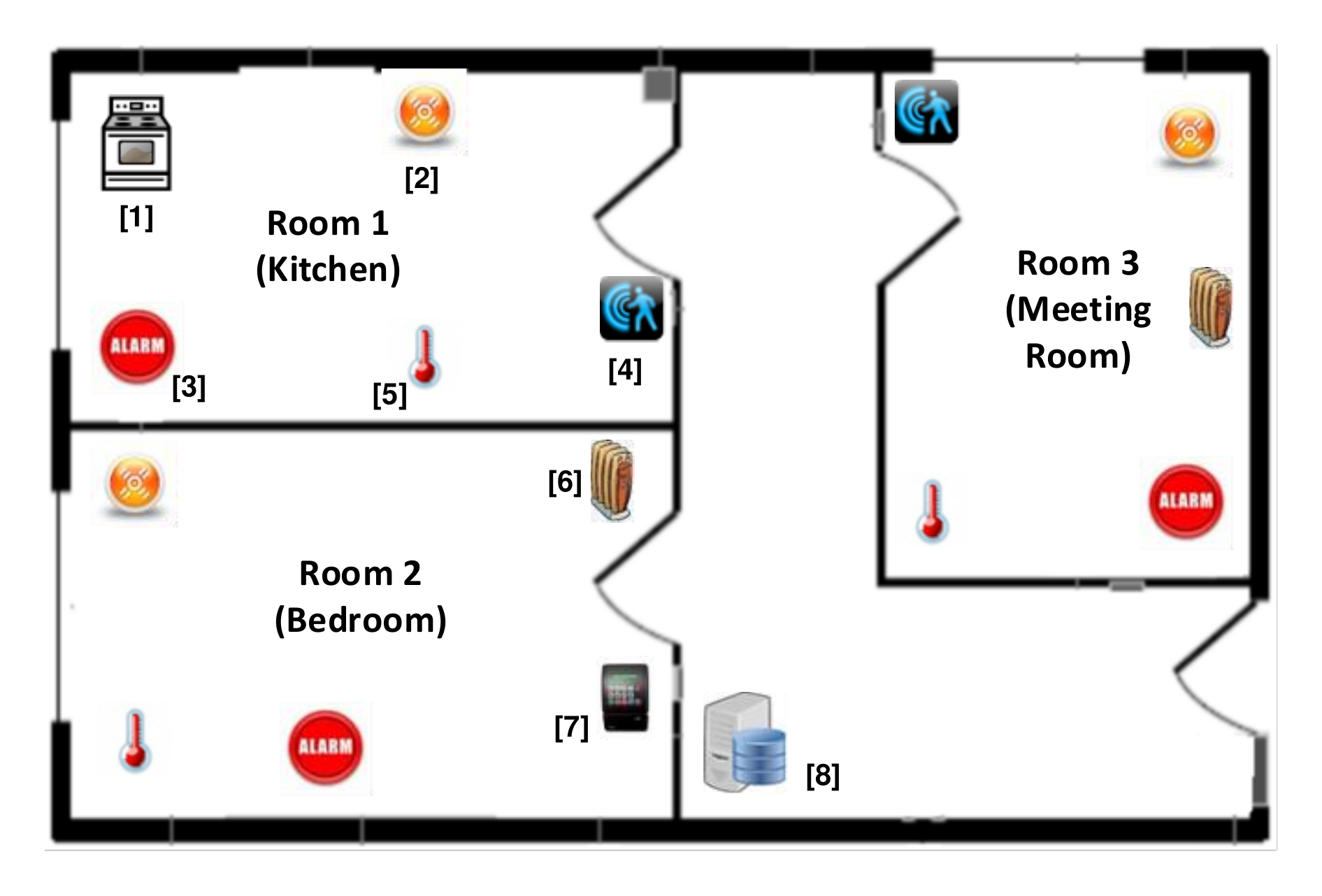}
\caption{A cluster of multi-floored buildings with deployed devices with [1]~stove, 
 [2]~smoke detector, [3]~alarm, [4]~motion sensor, [5]~temperature sensor, [6]~heater, [7]~badge reader, [8]~data storage.}
\label{fig:buildingautomationdomain}
\end{figure*}

\subsubsection{Evaluation}\label{sec:resultdevelopmenteffort}
We have implemented two IoT applications discussed in Section~\ref{sec:appdevelopmenteffort} 
using our approach. These applications are implemented independently. We did not reuse specifications 
and implementations of one application in other application. We deploy these two applications on simulated devices, 
running MQTT middleware that simulates network, on a single PC. 

\begin{figure*}[!ht]
\centering \includegraphics[width=0.7\textwidth]{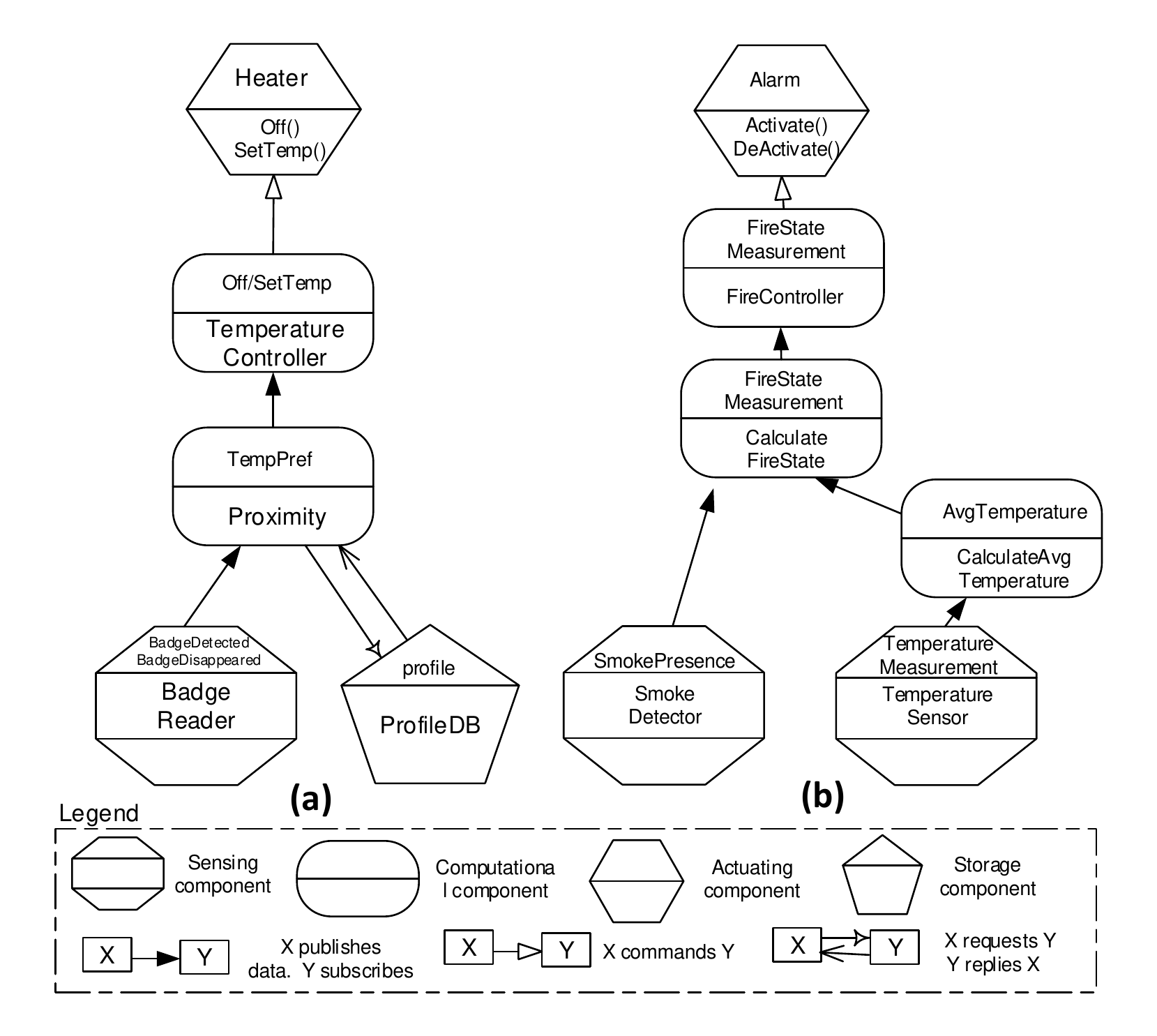}
\caption{Layered architecture of (a) \emph{personalized HVAC}  and (b) \emph{fire detection} application.}
\label{fig:personalizedHVACandFiredetection}
\end{figure*} 

We measured the \emph{lines of code} using Eclipse EclEmma 2.2.1 plug-in\footnote{\url{http://www.eclemma.org/}}.
This tool counts actual Java statement as lines of code and does not consider blank lines or lines 
with comments. Our measurements reveal that the percentage of handwritten lines of code, produced by stakeholders, 
is very low in both applications~(see Figure~\ref{table:locapp}). The measure of lines of code is only useful 
if the generated code is actually executed. We measured \emph{code coverage} of the generated 
programming frameworks~(i.e., mapping framework, vocabulary framework, architecture framework) 
of two applications~(see Figure~\ref{table:locapp}) using the EclEmma Eclipse plug-in. Our
measures show that more than 80\% of generated code is actually executed, the 
other portion being error-handling code for errors that did not happen during the experiment 
and/or unused features such as getter and setter. This high value indicates that most of the 
execution is spent in generated code and that, indeed, our approach reduces development effort by generating
useful code.

%a significant percentage of the total number of lines of code is generated in 
%two applications~(see Table~\ref{table:locapp}).  

\begin{figure*}[!ht]
\centering \includegraphics[width=0.70\textwidth]{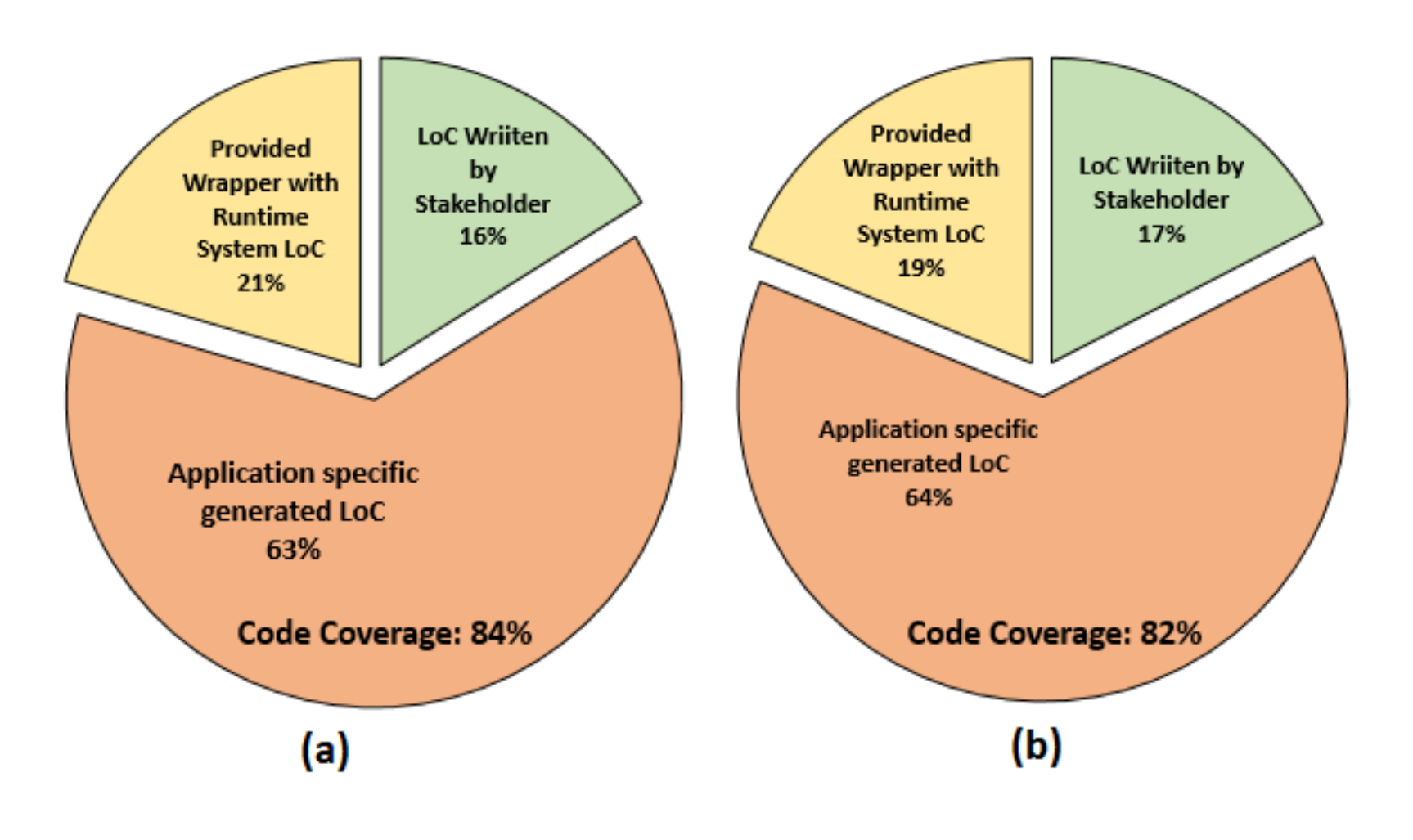}
\caption{Percentage of lines of code in (a) \emph{personalized HVAC} and (b) \emph{fire detection} application.}
\label{table:locapp}
\end{figure*}

\subsection{Development effort for a large number of devices}\label{sec:large}
The experiments, described in the previous sections, was conducted for 
a small number of devices. It does not demonstrate development effort for a large number of devices.  
Therefore, the primary aim of this section is to evaluate effort to develop an IoT application involving a large number of devices. 

In order to achieve the above aim, we have developed the \emph{road traffic monitoring
\& control application}~\cite{scopemottola2007}, depicted in Figure~\ref{fig:roadtrafficapplication}, 
that aims to maximize the flow of vehicles on the road. This kind of system is 
divided in disjoint sectors. Each sector is controlled depending on 
the current status of the sector.   In each sector, data is first collected from speed sensors 
and presence sensors and measurements such as average speed of vehicles and average 
queue length on a ramp are derived. The aggregated information is fed to an algorithm to determine the best actions 
to achieve the system objective -- maximize the flow of vehicles on the 
highway in each sector. The actions are then communicated to the ramp signals.

The application is developed on a set of simulated devices, running real
MQTT middleware, on a single PC. The assessments were conducted over an increasing number of devices.  
The first development effort assessment was conducted on 6 devices instrumented 
with sensors and actuators. In the next subsequent assessments,  we kept increasing 
the number of devices equipped with sensors and actuators.  In each 
assessment, we have measured lines of code to specify vocabulary, architecture, and deployment, 
application logic, and device drivers. Figure~\ref{fig:locsmartoffice} illustrates 
the assessment results containing a number of devices involved in the experiment and hand-written 
lines of code to develop the road traffic \& monitoring application.

\begin{figure*}[!ht]
\centering \includegraphics[width=0.70\textwidth]{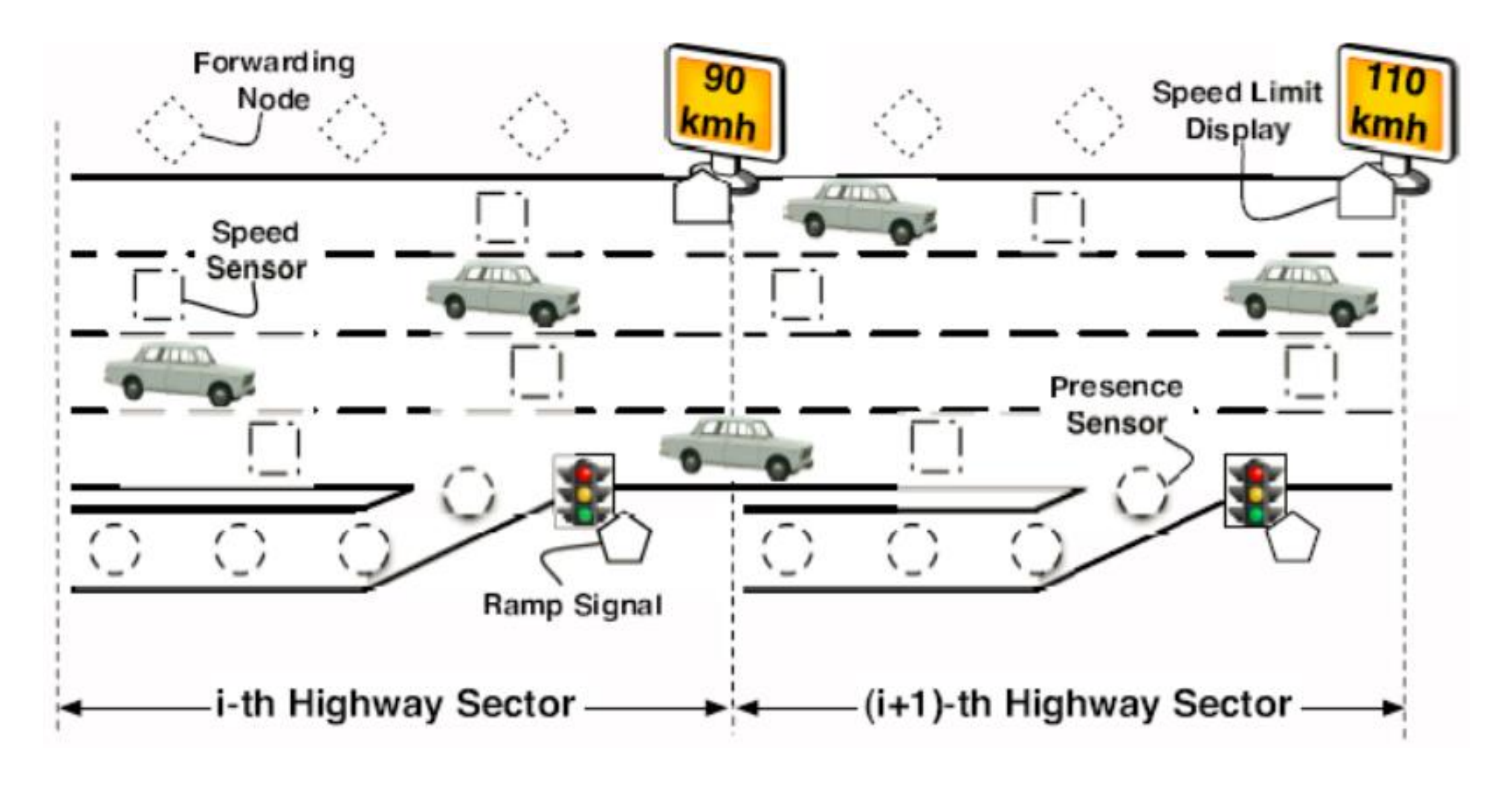}
\caption{Road traffic \& monitoring application scenario~(Image credit to \cite{scopemottola2007}).}
\label{fig:roadtrafficapplication}
\end{figure*}

\begin{figure*}[!ht]
\centering \includegraphics[width=0.60\textwidth]{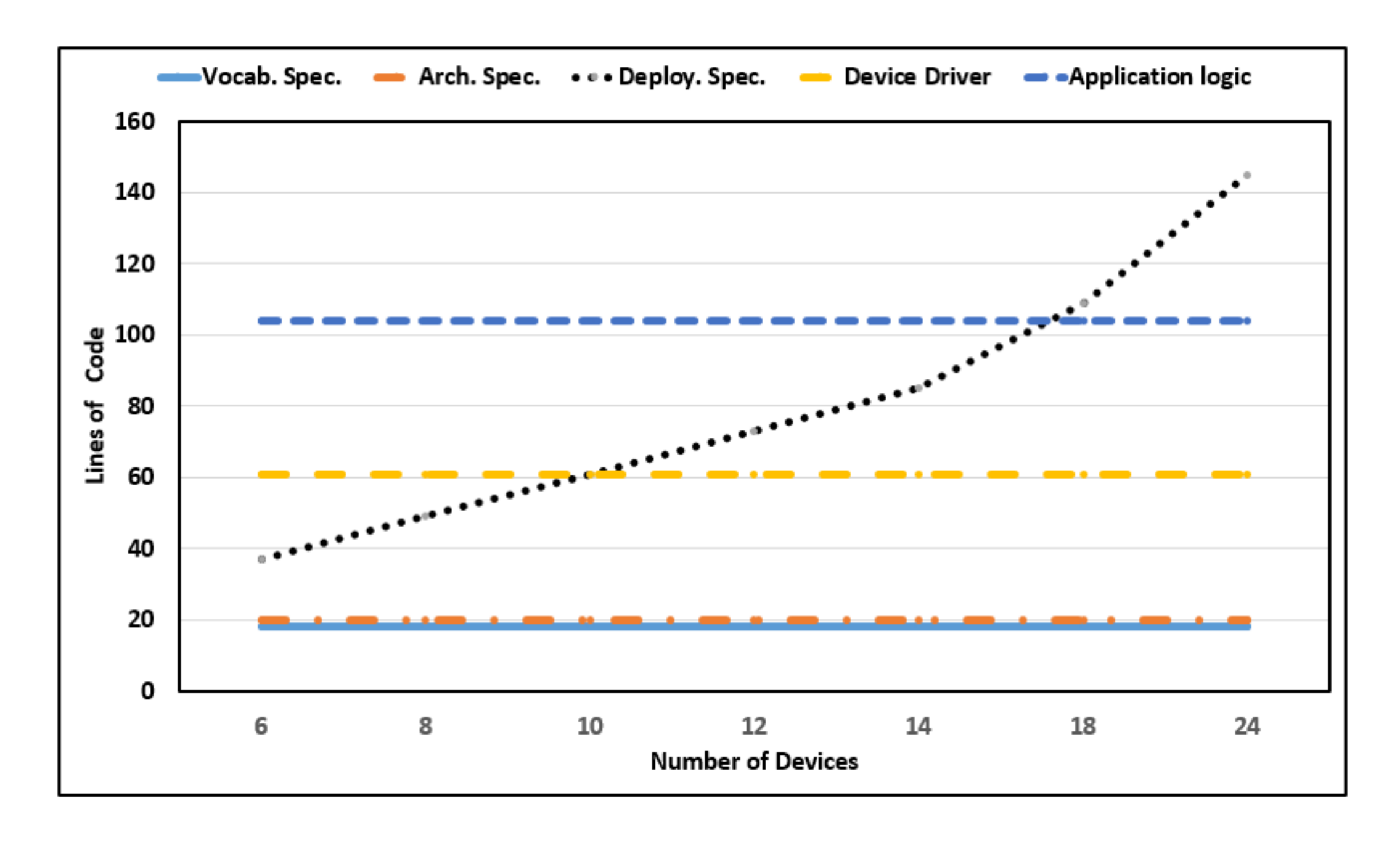}
\caption{Lines of code to develop the \emph{road traffic monitoring \& control} application.}
\label{fig:locsmartoffice}
\end{figure*}

In Figure~\ref{fig:locsmartoffice}, we have noted the following two observations and their reasons: 
\begin{itemize}
	\item As the number of devices increases, lines of code for \emph{vocabulary and architecture specification}, 
	\emph{device drivers}, and  \emph{application logic} remain constant, even for a deployment 
	consisting a large number of devices. The reason is that our approach provides 
	the ability to specify an application at a global level rather than individual nodes. 
	It means one entity description for many implementations and instances. Thus, development
	effort does not depend on the number entities.
	
	\item As the number of devices increases, lines of code for a deployment specification increase. The reason is that the 
			network manager specifies each device individually in the deployment specification. 
			This is a limitation of \emph{deployment language}. Our future work will  be to investigate 
			how a deployment specification  can be  expressed in a  concise and  flexible way for a network 
			with a large number  of devices. We believe that the use of  regular expressions 
			is a possible technique to address this problem. 

	\end{itemize}

\subsection{Reusability}\label{sec:reusability}
This section demonstrates the reusability of software artifacts using our approach. 
We consider two scenarios to demonstrate it: (1) evolving 
deployment~(detailed in Section~\ref{sec:changeintargetdeply}). 
(2) evolving  functionality~(detailed in Section~\ref{sec:changeinfunctionality}). 
The results of these experiments conclude that there is a drastic reduction in 
development effort for subsequent applications.

%(1) \emph{deploying an application to different deployments}~(detailed in 
%Section~\ref{sec:changeintargetdeply}). (2) \emph{evolving new functionality into existing 
%deployment}~(detailed in Section~\ref{sec:changeinfunctionality}). The results of these experiments 
%conclude that there is a drastic reduction in development effort for subsequent application development.

%\subsubsection{Deploying an application to different deployments}\label{sec:changeintargetdeply}
\subsubsection{Evolving deployment}\label{sec:changeintargetdeply}
This scenario demonstrates the reusability of  specifications and implementations when the same application 
is deployed to different deployment scenarios. To illustrate it, we consider the home 
scenario shown in Figure~\ref{fig:buildingautomationdomain} and take the fire detection 
application~(discussed in Section~\ref{sec:appdevelopmenteffort}). The application 
is initially deployed in the \texttt{bedroom}. Latter for the safety reason, it is
decided to deploy the same application in the \texttt{kitchen} and \texttt{meeting room}.
Figure~\ref{fig:buildingautomationdomain} shows both the \texttt{kitchen} 
and \texttt{meeting room} have all necessary sensors and actuators to deploy 
the fire detection application.

%As shown in Figure~\ref{fig:buildingautomationdomain}, both the \texttt{kitchen} 
%and \texttt{meeting room} have all necessary sensors and actuators to deploy 
%the fire detection application.

%Now, an owner wants to deploy the same application 
%to the \texttt{kitchen} and \texttt{meeting room} for the additional safety. 

To measure the effort for developing the same fire management application for different deployment scenarios, 
we have simulated it on a set of simulated devices, running MQTT 
middleware that simulates network, on a single PC. Initially, when the application is developed using our 
approach in the~\texttt{bedroom}, we have written the vocabulary, deployment, and architecture specification, 
device drivers, and application logic from scratch. Table~\ref{table:developmenteffortDeployment} 
shows the lines of code for developing the fire detection application in the \texttt{bedroom}.
However, the reusability of specifications and implementations become apparent when 
we deploy it in the \texttt{kitchen} and \texttt{meeting room}. 
To develop subsequent applications, we only need to specify  deployment specification and can reuse 
other software artifacts. Table~\ref{table:developmenteffortDeployment} shows the drastic reduction 
in development effort for the \texttt{kitchen} and \texttt{meeting room}. We conclude that the primary reason of 
drastic reduction of development effort in the next two deployment scenarios using our approach 
is \emph{separation of concerns}.  Our approach separates the IoT application development 
into well-defined concerns. Therefore, stakeholders achieve high reusability of  specifications 
and implementations across applications of a same application domain. Thus, it  reduces the 
development effort.

\begin{table*}[!ht]
\centering 
\begin{tabular}{ p{6cm} | >{\centering\arraybackslash}p{0.7cm}>{\centering\arraybackslash}p{0.7cm}>{\centering\arraybackslash}p{0.7cm} >{\centering\arraybackslash}p{0.7cm}>{\centering\arraybackslash}p{0.7cm} }
 \toprule 

Application &	Vocab Spec. &	Arch. Spec. & Deploy. Spec. 
& Device driver & App. logic   \\ \midrule
Fire detection (in \texttt{bedroom})   & 20 & 16&49&101&60 \\ \midrule
Fire detection (in \texttt{kitchen})     & 0 & 0&49&0&0 \\ \midrule
Fire detection (in \texttt{meeting room}) & 0 & 0&49&0&0 \\
\bottomrule
\end{tabular}
\caption{The lines of code to develop the fire detection application. Initially, it was 
deployed in the \texttt{bedroom} and latter it was deployed to the \texttt{kitchen} and \texttt{meeting room}.} 
\label{table:developmenteffortDeployment} 
\end{table*}

\begin{figure*}[!ht]
\centering \includegraphics[width=0.80\textwidth]{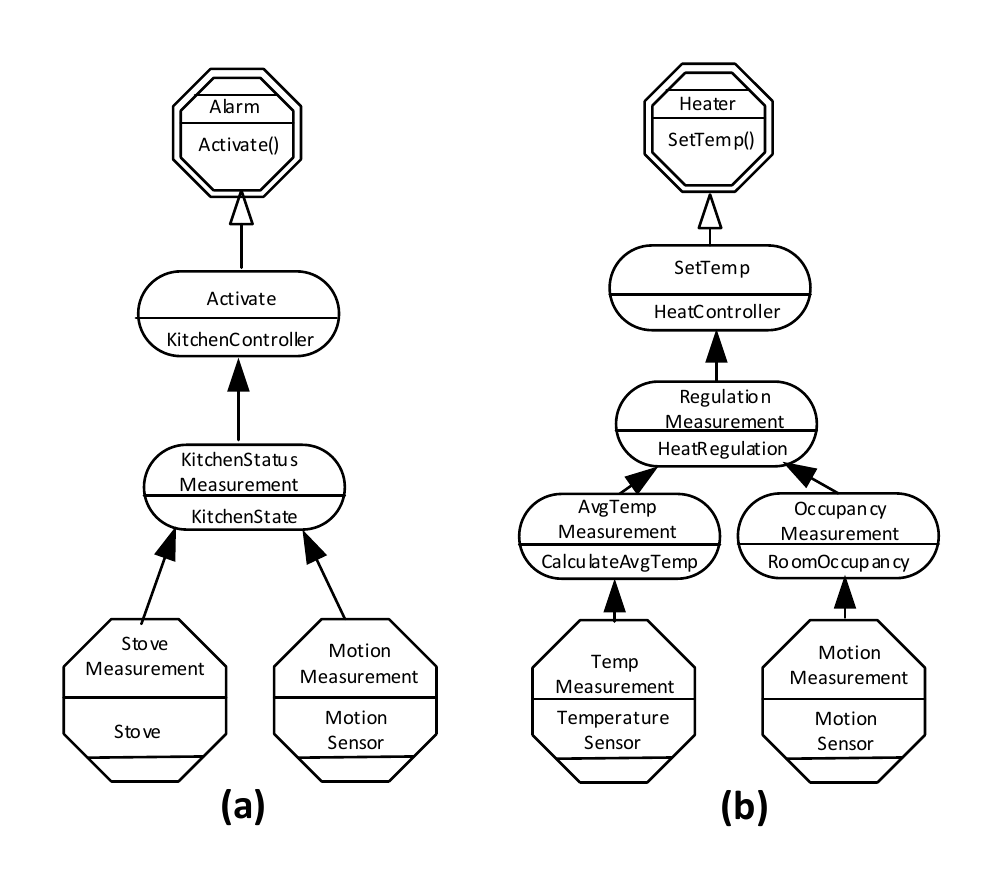}
\caption{Layered architecture of (a) safety in kitchen application and (b) heating control system.}
\label{fig:safetyandheatingcontrol}
\end{figure*} 

%when there
%is requirements of adding new functionality into existing deployment.

\subsubsection{Evolving functionality}\label{sec:changeinfunctionality}
This section demonstrates the reusability of specifications and implementations 
when functionality is evolved. To illustrate this scenario, we consider 
the home scenario shown in Figure~\ref{fig:buildingautomationdomain}. 
Apart from two applications, described in Section~\ref{sec:appdevelopmenteffort},
we consider a scenario of implementing the following two new functionality:

\begin{itemize}
	\item \emph{Safety in kitchen}~\cite{bischoff2008engineering} raises an alarm if the stove is switched on and 
when nobody in the kitchen. Figure~\ref{fig:safetyandheatingcontrol} shows a layered architecture of it.
The motion sensor detects the presence of moving objects. The smart stove senses if it is turned 
on or off. Based on these inputs, the system takes appropriate decisions.

\item \emph{Heating control system}~\cite{bruneau2012developing} regulates the temperature 
in the meeting room depending on the room occupancy. Figure~\ref{fig:safetyandheatingcontrol} 
shows a layered architecture of it. The system receives motion detection events to detect the occupancy. 
Thus, if a person enters into the meeting room and the average temperature is below the certain threshold, 
the heating control system automatically controls the temperature.

\end{itemize}

We have simulated the above applications on a set of simulated devices, running iBICOOP 
that simulates network, on a single PC. Initially, when the \emph{personalized HVAC},  
\emph{fire detection}, \emph{safety in kitchen} applications are developed using our
approach, we have written vocabulary specification, architecture specification, deployment
specification, application logic, and device driver. The labels \texttt{(a1)}, \texttt{(b1)}, 
\texttt{(c1)}  in Figure~\ref{table:developmenteffortDeploymentinEvolution} 
show the lines of code for developing these three applications and the corresponding labels \texttt{(a2)}, 
\texttt{(b2)}, and \texttt{(c2)} describe the components specified in the vocabulary specification.

The reusability of previously written  components becomes apparent when we develop 
the \emph{heating control} application. The components specified in the previous three 
applications are reused to write vocabulary specification and the device driver. 
The label \texttt{(d2)} in Figure~\ref{table:developmenteffortDeploymentinEvolution} 
indicates the reusability of the previously written temperature sensor, heater, and motion sensor components with
similar patterns and colors. The \texttt{(d1)} in Figure~\ref{table:developmenteffortDeploymentinEvolution} 
shows the drastic reduction in the lines of code for the heating control application. More
specifically, the lines of code for the vocabulary specification and device driver remains zero.
We conclude that the primary reason of drastic reduction of development effort for the heating 
control system using our approach is \emph{separation of concerns}.  Since, our approach 
separates the IoT application development into well-defined concerns. Therefore, stakeholders achieve high 
reusability of  specifications and implementations across applications. 
Thus, it reduces the development effort.

\begin{figure*}[!ht]
\centering \includegraphics[width=0.90\textwidth]{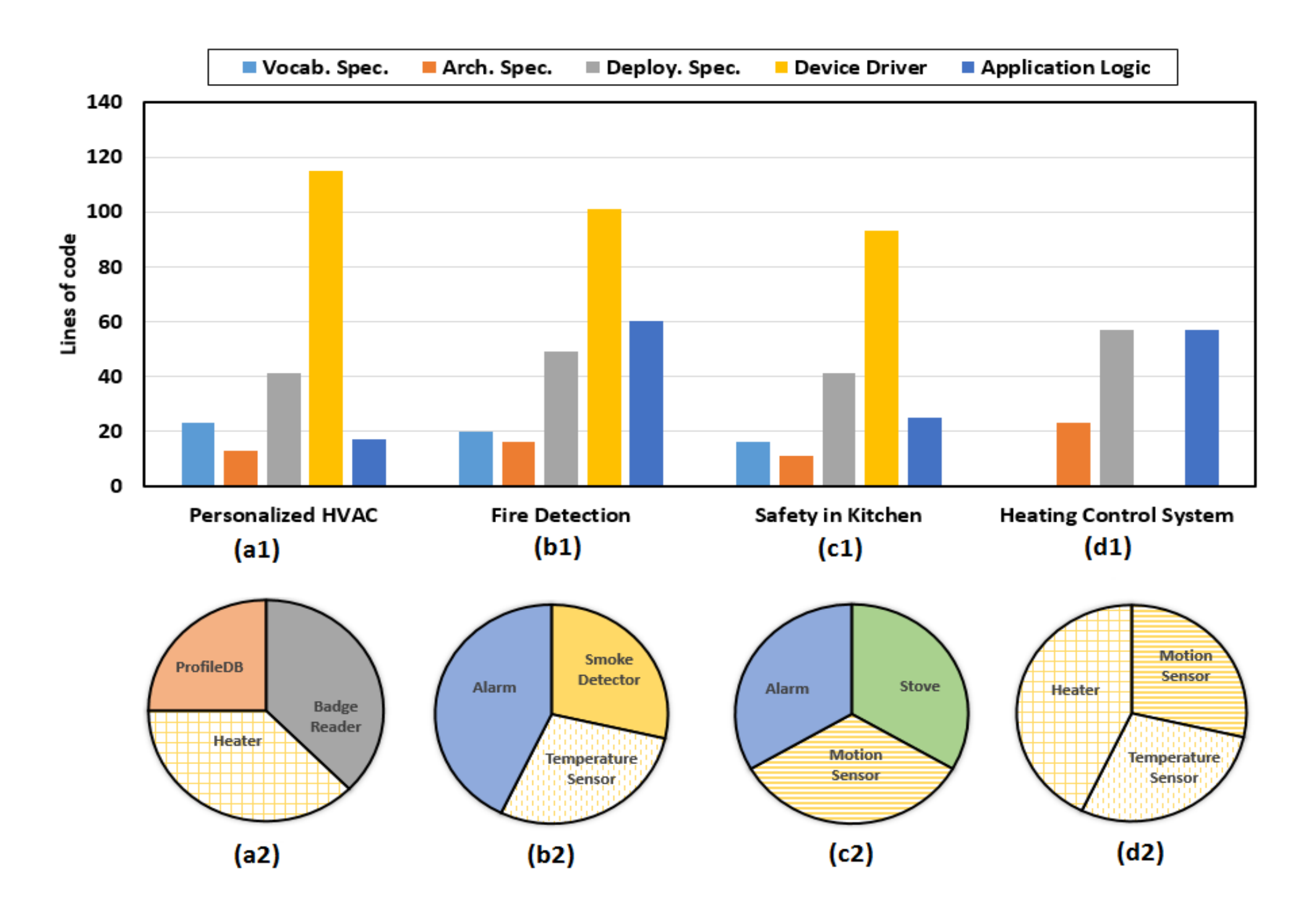}
\caption{ The labels \texttt{(a1)}, \texttt{(b1)}, \texttt{(c1)}, and \texttt{(d1)}  
show the lines of code for developing the personalized HVAC, fire detection, safety kitchen, 
and heating control system applications and the corresponding labels \texttt{(a2)}, \texttt{(b2)}, 
\texttt{(c2)}, and \texttt{(d2)} describe the components specified in the vocabulary specification.}
\label{table:developmenteffortDeploymentinEvolution}
\end{figure*}

\subsection{Expressiveness}\label{sec:expressiveness}
This section evaluates the scope of our approach. More specifically, it answers the question:
\emph{What are the characteristics of IoT applications that can be modeled by our approach?}
In order to answer this question, we map the IoT applications~(discussed in
Section~\ref{sec:appexample},\ref{sec:developmenteffort},\ref{sec:large}) into identified 
characteristics using our approach. Table~\ref{expressiveness} 
indicates the subset of IoT application characteristics that can be modeled 
using our development framework. It notes the following observations about our approach.

\fakeparagraph{\emph{Heterogeneous components}} An IoT application may execute on a network consisting of 
different types of components. For example, the smart building application consists of components, including
sensing (e.g., temperature sensor, badge reader), actuating (e.g., heater, light), storage (e.g., proﬁle storage on
different database systems such as MySQL or MongoDB). As indicated in Table~\ref{expressiveness}, our approach 
supports components commonly found in IoT applications .

\fakeparagraph{\emph{Heterogeneous platforms}} An IoT application may execute on a network with 
heterogeneous platforms. These platforms are operating system-specific. For instance, a device could be
running Android mobile OS, Java SE on laptops, or a server OS such as GNU/Linux etc.  
Table~\ref{expressiveness} shows the supported platforms in the current version. It generates code 
for JavaSE and Android platforms. However, our approach is flexible to generate code 
for different platforms, as discussed in our previous work~\cite{soukaras2015iotsuite}.

\fakeparagraph{\emph{Heterogeneous runtime system}} An IoT application may consist of devices, running 
heterogeneous runtime system that are responsible for a distributed execution of an application. 
Table~\ref{expressiveness} shows the supported runtime systems in the current version. It implements
the MQTT and iBICOOP runtime system. However, the framework does not restrict the stakeholders 
to a specific runtime systems and it is flexible to integrate different runtime systems, 
as discussed in our previous work~\cite{soukaras2015iotsuite}.

\fakeparagraph{\emph{Heterogeneous interaction modes}} The devices could be different in terms 
of how data can be accessed from them. Table~\ref{expressiveness} shows four interactions
modes supported by our approach, largely found in the IoT applications: periodic~\cite{programmingWSNmottola2011}, publish/subscribe~\cite{eugster2003many}, request/response~\cite{berson1996client}, 
command~\cite{andrews1991paradigms}.

\fakeparagraph{\emph{Topology}}
It indicates whether an application is characterized by static or dynamic topology.
In static topology, devices do not move once they are deployed. In dynamic topology, 
devices (e.g., smart phone) move autonomously. Table~\ref{expressiveness} shows 
that applications with static topology is supported.  Our immediate future work will be to 
provide mobility support in our framework.

\fakeparagraph{\emph{Network size}}
It indicates a number of devices participating in an application~\cite{designspaceromer2004}. 
The participating devices could be sensing, actuating, computational, and/or user interface devices.  
Table~\ref{expressiveness} shows our approach can cover an application from a small 
to large number of devices.

\fakeparagraph{\emph{Behaviors in IoT applications}} To guide us our efforts in creating a development framework 
for the IoT, we survey various applications present in the research literature as well as commercial product 
proposals. Our early study came to the conclusion that the IoT brings the following behaviors 
to the mix (refer our previous work~\cite{appdevIoTpatel2011} for more detail).
Table~\ref{expressiveness} shows the behaviors of IoT applications that are covered by the development framework.

\begin{itemize}
	\item 
	\fakeparagraph{\textbf{Intermittent sensing}} This behavior comes from the early definition 
	of the IoT, which was centered around RFID technology, and is found mostly in applications where 
	things have an {\em information shadow}~\cite{29} on the Internet. The reader (e.g. RFID reader, 
	barcode reader) observes an ID of a tag and sends it to a service on the Internet, which fetches 
	data associated with the ID from the storage and returns it to the application.

\item 
	\fakeparagraph{\textbf{Regular data collection}} This behavior is seen in the class of IoT applications 
	where (smart) things interact with users by stating information about themselves \emph{periodically} or \emph{event-based}. 
	Actual objects are observed by sensors, and then the observed information is sent for various purposes 
	to users. One of classic examples could be found in the building automation domain. One of ways to save energy is 
	to engage the residents of the building. A system can generate situation awareness 
	by displaying general information about the building such as current temperature or energy 
	usage\footnote{\url{http://www.arrayent.com/}}   of the 	building on dashboard placed on 
	a central location of a building.  

\item \fakeparagraph{\textbf{Sense-Compute-Actuate loops}} This behavior is seen in applications where 
smart things interact with each other at either the local level or through the Internet, and provide 
information that can be used as new knowledge\footnote{\url{http://www.casaleggio.it/internet_of_things/}}. 
They may also take corrective actions~\cite{3} with no human originator, recipient, or intermediary, 
and may notify or prompt users as required.
\end{itemize}

\subsection{Memory metrics}\label{sec:codemetrics}
Increasing stakeholders’ productivity often comes at a cost~\cite{sivieri2014eliot}.  To precisely evaluate 
this aspect, we measure the average  \emph{code size} and \emph{memory consumption} of device-specific packages 
that are deployed on devices. These two metrics are important because they give approximate indication 
of the amount of memory a device needs to run an application. A device with less memory 
is not able to run the application implementation.

\newcolumntype{A}{>{\columncolor[gray]{0.8}[\tabcolsep][\tabcolsep]}c}
\newcolumntype{B}{>{\columncolor[gray]{0.95}} c|}
\begin{table*}[!ht]
\centering
\begin{scriptsize}
\renewcommand{\arraystretch}{1.5} 
\onecolumn
\begin{longtable}{
>{\centering\arraybackslash}m{2.5cm}
>{\centering\arraybackslash}m{1.7cm}
>{\centering\arraybackslash}m{1.2cm}
>{\centering\arraybackslash}m{1.6cm} 
>{\centering\arraybackslash}m{1.0cm}  
>{\centering\arraybackslash}m{0.8cm} 
>{\centering\arraybackslash}m{2.0cm} 
%>{\centering\arraybackslash}m{1.0cm} 
>{\centering\arraybackslash}m{1.0cm} 
>{\centering\arraybackslash}m{1.0cm} 
} \toprule
Application Domain& Application Name & Behaviors & Components & Platform&Runtime System &Interaction Modes& Topology & Network size\\ \toprule 
\multirow{4}{*}{Building Automation} & Personalized HVAC& SCC loop  & Sensor, Storage, Computation, Actuator& JavaSE, Android & MQTT &Event-driven, Request-Response, Command & Static & 5 \\ \cmidrule{2-9}
& Fire Detection & SCC loop & Sensor, Computation, Actuator & JavaSE & MQTT &Event-driven, Periodic, Command & Static & 8 \\ \cmidrule{2-9}
 & Safety in Kitchen& SCC loop  & Sensor, Computation, Actuator & JavaSE, Android & iBICOOP &Event-driven, Periodic, Command & Static & 5 \\ \cmidrule{2-9}
& Collecting Avg Temperature of a building & Regular data collection & Sensor, Computation, User interface 
& JavaSE & MQTT  & Periodic, Command & Static & 16 \\ \midrule
Traffic Management& Road Traffic Monitoring \& Control& SCC loop & Sensor, Computation, Actuator & JavaSE & MQTT &Event-driven, Periodic, Command & Static & 24 \\ 
\toprule
\endfirsthead
\\
\caption{Expressiveness of our approach}
\label{expressiveness}
\end{longtable}
\twocolumn
\end{scriptsize}
\end{table*}

\fakeparagraph{\emph{Code size}} IoT devices possess the limited amount of program memory. 
So, the code size is an important metric to measure for IoT applications~\cite{oopluca2014}.

\fakeparagraph{\emph{Memory consumption}} RAM is a severely limited resource in devices. So, 
it is important to be know memory consumption of a device. We measure the average amount of heap space used by  
device-specific packages using VisualVM\footnote{\url{https://visualvm.java.net/}}: a tool that  
reports the heap allocated of a running application. To get precise results, we separately 
asses the virtual machine with each package code loaded and when the whole application running. 

Table~\ref{table:memorymetrics} shows an average code size and heap allocated for the applications. 
These results give not very precise, but approximate, indicate  the amount of memory a 
device needs to run an application, thus guiding stakeholders to choose 
an appropriate device run to device-specific code generated as final package using our approach.
\begin{table*}[!ht]
\centering \scriptsize
\begin{tabular}{lp{3.7cm}p{3.7cm}}
 \toprule
\textbf{Application}  & \textbf{Average code size~(in MBs)} & \textbf{Average memory consumptions~(in MBs)} \\ 
 \midrule
Personalized HVAC      & 18.58 & 16.95 \\  \midrule
Fire detection         & 18.40  & 16.91 \\ 
\bottomrule
\end{tabular}
\caption{Average code size and memory consumptions of a device-specific code} 
\label{table:memorymetrics} 
\end{table*}

\subsection{Comparison with state of the art}\label{sec:comparison}
This section takes a snapshot of the current approaches available for the IoT application development 
and compares them with our approach. We begin by presenting various dimensions that characterize 
application development approaches. Then, we map existing  approaches 
back to the dimensions in Table~\ref{table:IoTApplicationDevApproaches}, therefore providing not 
only a complete view of the state of the art with respect to our approach, but also useful 
insights for selecting the most appropriate approach given an application requirement at hand. 
Table~\ref{table:IoTApplicationDevApproaches} shows the 
comparative analysis with our approach. Due to similarity, we pick one representative 
system in some cases. For instance, TinyDB and Cougar have adopted SQL-based interface 
for collecting data. So, we take only TinyDB as a representative example.   

\fakeparagraph{\emph{Classification}} We see application development  approaches for the IoT are classified into
four broad categories: \emph{node-centric programming}, \emph{database approach}, \emph{macro-programming}, 
and \emph{model-driven approach}. For detail descriptions with pros and cons, readers are referred 
to our work~\cite[p.~9]{patel-thesis14}. 

%\subsubsection{Dimensions}\label{sec:dimension}

%The goal of this section is to compare the existing approaches with our approach 
%in a quantitative way. For that, we have established our own set of dimensions. The set is not exhaustive. 
%However, it provides a guide to stakeholders for selecting an appropriate approach. 
%The set is composed of the following dimensions:

\fakeparagraph{\emph{Systematization of the development process}}
It defines a precise sequence of steps to be followed to develop IoT applications, 
as well as identifies roles of each stakeholder and separates them according 
to their skills. The clear identification of expectations and specialized skills 
of each type of stakeholders helps them to play their part effectively. 
Hence, this separation of roles smoothen the application development process

\fakeparagraph{\emph{Technological change support}} It indicates the support provided by an approach 
to integrate runtime systems and programming languages. The key advantage of this feature it gives a flexibility to extend 
an approach with a new supported communication technologies and new programming language.

\fakeparagraph{\emph{Programming interface}} It indicates interface provided by an approach 
to programmer in order to specify an application.  An approach largely provides Domain-specific Language~(DSL), 
General-purpose Programming Language~(GPL), or combinations of both. \emph{DSL} could be graphical/textual. 
Examples of graphical DSL are drag and drop blocks or UML notations. Examples of GPL are Java, C, etc. 

\fakeparagraph{\emph{Open source}} It indicates whether an approach is open source or not. Given 
the usefulness of open source, it aims to provide an opportunity of 
sharing of novel software engineering tools and technologies. 

\fakeparagraph{\emph{Integrated Development Environment~(IDE) Support}}  
It indicates an IDE is provided by an approach or not. An IDE is a software 
application that provides comprehensive facilities for software 
development. It normally consists of editors, which facilitate syntax coloring and error 
reporting, and automation tools to reduce application development effort.

\fakeparagraph{\emph{Deployment state}}  It represents maturity of an approach. Whether it is just a prototype 
or it has been released as product.

\begin{table*}[!ht]
\centering 
\tiny
\begin{tabular}
{p{1.2cm}>{\centering\arraybackslash}p{2.0cm}>{\centering\arraybackslash}p{0.5cm}>{\centering\arraybackslash}
p{2.5cm}>{\centering\arraybackslash}p{2.5cm}>{\centering\arraybackslash}p{0.4cm}>{\centering\arraybackslash}
p{0.4cm}>{\centering\arraybackslash}p{1.0cm}>{\centering\arraybackslash}p{0.5cm}>{\centering\arraybackslash}
p{0.5cm}>{\centering\arraybackslash}p{0.5cm}>{\centering\arraybackslash}p{0.5cm}>{\centering\arraybackslash}
p{0.2cm}
}
\toprule
 &  & &  &	 &  &   &  &     \multicolumn{5}{c}{Automation at Development Life cycle} \\ \cmidrule{9-13}
Classification & Approaches & Sys. of dev. Proc. & Programming Interface  & IDE Support & Open Source & Deply. State
& Tech. Change Support  & Req. Analysis	& Design & Impl.	& Testing	& Deply. \\ \midrule

&Android Wear~(2014) & N  & Java  & Eclipse plug-in, Android Studio& PL& S & N & N & N & Y & Y & Y \\\cmidrule{2-13}
&Apple HomeKit~(2014)& N  & Objective-C  & XCode  & N  & S & N & N & N & Y & Y & Y\\\cmidrule{2-13}
Node-centric &RedHat JBOSS A-MQ(2012)& N  &Java/C/C++ & JBoss Developer  & Y  & S & N & N & N & Y & Y & Y \\\cmidrule{2-13}
Programming  & DeviceHive~(2013) & N  &Java/C++/.Net/ JavaScript/Python  & Not clear  & Y  & S & N & N & N & Y & N & N\\\cmidrule{2-13}
& Electric imp~(2011) & N  & Squirrel  & ElectricIMP IDE  & N  & S & N & N & N & Y & Y & Y   \\\cmidrule{2-13}
& WSO2(2005)               & N  & Jaggery.js  & WSO2 Developer Studio  & Y  & S & N & N & N & Y & Y & Y\\ \cmidrule{2-13}
& Samsung SmartThings~(2012) & N  & Groovy  & Web-based IDE  & N  & S & N & N & N & Y & Y & Y \\\midrule
& Dominique et al.~\cite{guinard2010resource}(2010) & N  & DSL~(Graphical)  & Web-based IDE  & Y & P & N & N & Y & N & N & Y   
\\\cmidrule{2-13}
& Intel IoT~(2014) & N  & DSL~(Graphical)\&C/C++  & Eclipse plug-in, Wyliodrin  & N  & S & N & N & Y & Y & Y & Y\\\cmidrule{2-13}
Macro& Microsoft HomeOS~\cite{MicrosoftHOMEOS2012}(2012) & N  & DSL~(Graphical)  & Visual Studio  & Y  & S & Y & N & Y & Y & Y & Y\\\cmidrule{2-13}
Programming& MuleSoft Anypoint~(2013) & N & DSL~(Graphical)\&RAML/XML  & Web-based IDE, Anypoint Studio& N  & S & N & N & Y & Y & Y & Y \\ \cmidrule{2-13}
& Open IoT~\cite{OpenIoT2014}(2014) & N  & DSL~(Graphical)  & Eclipse plug-in  & Y  & P & N & N & Y & Y & Y & Y\\ \midrule
& TinyDB~\cite{madden2005tinydb}(2005) & N  & SQL-like & Not clear  & Y & S & N & N & NA & Y & NA & NA\\ \cmidrule{2-13}
Database          
&TinySOA~(2009)& N  & Service-oriented APIs&TinyVisor&Y&P&Y&N&N&Y&NA&NA\\\cmidrule{2-13}
Approaches	& Priyantha et al.~\cite{priyantha2008tiny}(2008) & N  & SOAP-based APIs & Visual Studio, NetBeans & N  & P & N & N & N & Y & NA & NA\\  \midrule
& \textbf{\texttt{Our approach~(2014)}} & \textbf{\texttt{Y}}& \textbf{\texttt{DSL~(Textual)}} \& \textbf{\texttt{Java}  }& \textbf{\texttt{Eclipse plug-in}}  & \textbf{\texttt{Y}}  & \textbf{\texttt{P}} & \textbf{\texttt{Y}} & \textbf{\texttt{N}} & \textbf{\texttt{Y}} & \textbf{\texttt{Y} }& \textbf{\texttt{N}} & \textbf{\texttt{Y}}   \\ \cmidrule{2-13}
Model-driven & DiaSuite~\cite{towardscassou2011}(2011) & Y  & DSL~(Textual) \& Java  & Eclipse plug-in  & N  & P & Y & N & Y & Y & Y & N \\\cmidrule{2-13}
Approaches & Srijan~\cite{pathak2007compilation}(2008) & N  & DSL~(Graphical) \& Java  & Eclipse plug-in  & Y  & P & N & N & Y & Y & N & Y\\\cmidrule{2-13}
& PervML~\cite{pervMLserral2010}(2009)  & Y & DSL~(Graphical)  & Eclipse plug-in  & N  & P & Y & N & Y & N & Y & N\\\cmidrule{2-13}
& RuleCaster~\cite{bischoff2006rulecaster}(2007) & N  & DSL~(Textual) & Command Line  & N & P & N & N & Y & N & N & Y \\ \bottomrule
\end{tabular}
\caption{Comparison of existing approaches with our approach. (Deply.-Deployment, Impl.-Implementation,
Req.-Requirement, Tech.-Technological,  H-High, L-Low, M-Medium, P-Prototype, S-Stable, PL-Partial, NA-Not Applicable, Y-Yes, N-No, Sys.- Systematization, Dev.-Development, Pro.-Process).}
\label{table:IoTApplicationDevApproaches}
\end{table*}

\section{Conclusion}\label{chapt:conclusionandfuturework}
An important challenge that needs to be addressed in the IoT is 
to enable the IoT application development with minimal effort by the various 
stakeholders. To address this challenge, many approaches have been proposed.
Although existing approaches provide a wide range of features, stakeholders have 
specific application development requirements and choosing an appropriate approach 
requires thorough evaluations on different aspects. 

This paper is an attempt to address the above issue partially.
It presents a set of evaluations based on our previous work on IoT application development framework.  
This paper evaluate our approach in terms of \emph{development effort}, \emph{reusability}, \emph{expressiveness}, 
\emph{memory metrics}, and \emph{comparison with the state of the art} in IoT application 
development one various \emph{dimensions}.  The set of measures is non-exhaustive. However, they 
reflect principal quantitative advantages that our approach provides to stakeholders.  
Moreover, these measures provide the research community with insight into evaluating, 
selecting, and developing useful IoT frameworks and applications. 

\balance
\bibliographystyle{abbrv}
%\bibliography{references} 

\end{document}